\documentstyle[preprint,aps]{revtex}
\preprint{SNUTP 97-003}

\begin{document}
\draft
\title{
Quantum theory of motion of a time-dependent harmonic oscillator
in the pilot-wave theory
}

\author{
Jeong-Young Ji\footnote{
Electronic address: jyji@phyb.snu.ac.kr}
and
Kwang-Sup Soh\footnote{
Electronic address: kssoh@phyb.snu.ac.kr}
}
\address{Department of Physics Education,
Seoul National University, Seoul, 151-742, Korea}
\maketitle

\begin{abstract}
The de Broglie-Bohm quantum trajectories are found in analytically closed 
forms for the eigenstates and the coherent state of the Lewis-Riesenfeld 
(LR) invariant of a time-dependent harmonic oscillator. It is also shown 
that an eigenstate (a coherent state) of an invariant can be interpreted as 
squeezed states obtained by squeezing an eigenstate (a coherent state) of 
another invariant. This provides ways for a whole description of squeezed 
states. 
\end{abstract}

\newpage

The quantum theory of motion introduced by de-Broglie and Bohm provides 
another viewpoints on the world of quantum mechanics~\cite{deBroglie,Bohm}. 
There are many examples which explain the structure of the atom, the 
interference and the tunneling in the quantum motional 
scheme~\cite{Holland}. 
The quantum trajectories have been numerically solved in a time-dependent 
scattering from square barriers and square potential wells~\cite{Dewd,Hirs}. 
Recently, the damped harmonic oscillator is analyzed according to de 
Broglie-Bohm theory~\cite{Vand}. 
However explicitly time-dependent systems have not yet been dealt with. 
Since the exact wave function for an explicitly time-dependent harmonic 
oscillator is recently found~\cite{Exac}, we expect that one may calculate 
the quantum trajectories for an eigenstate, a coherent state, and a squeezed 
state of the system.

In this Letter, we find the de Broglie-Bohm quantum trajectories for a 
time-dependent harmonic oscillator. The LR invariants~\cite{LR} are 
constructed using the creation and the annihilation operators which are 
expressed in terms of a classical solution. It is argued that since any 
linear combination of two independent classical solutions is also a 
solution, there are many invariants and the invariant can be specified 
uniquely (except for a constant factor) if we fix the classical solution. It 
is shown that the invariants with different solutions are obtained by 
squeezing each other with a squeezing operator. Here the squeezing operator 
describes the Bogoliubov transformation between two sets of creation and 
annihilation operators. For an invariant, we consider the eigenstates and 
coherent states as a guiding wave in the de Broglie-Bohm theory and find the 
corresponding quantum trajectories. Because these states are squeezed states 
for another invariant, we have a whole description of squeezed states. The 
solutions are applied to a time-independent oscillator and a damped 
oscillator. In the damped oscillator we find the new type of quantum 
trajectories which have not been found in Ref.~\cite{Vand}. Here the quantum 
motions are oscillating while damping.

In the causal interpretation of quantum mechanics, the wave is 
mathematically described by $\Psi ({\bf x},t) $, a solution to 
Schr\"odinger's wave equation

\begin{equation}
i \hbar \frac{{\partial \Psi}}{\partial t} = 
\left( - \frac{{\hbar^{2}}}{2m} \nabla^{2} + V \right)
\Psi
\end{equation}
Setting $\Psi = R e^{iS/\hbar} $ where $R $ and $S $ are real functions of position and time, the particle trajectories of a 
particle of mass $m$ moving in the potential $V+Q$ is described by the 
following quantum Hamilton-Jacobi equation:
\begin{equation}
\frac{{\partial S}}{\partial t} + \frac{{(\nabla S)^{2}}}{2m} 
 + V + Q = 0 
\end{equation}
where the quantum potential is given by
\begin{equation}
Q = - \frac{{\hbar^{2}}}{2m} \frac{{\nabla^{2} R}}{R} 
\end{equation}
Thus the particle trajectories associated with a given quantum mechanical 
state can be obtained by solving the quantum equation of motion
\begin{equation}
\dot{\bf x} = (1/m) \nabla S ( {\bf x}, t) |_{{\bf x}={\bf x}(t)}
\label{QEOM}
\end{equation}
or equivalently
\begin{equation}
m \ddot{\bf x} = - \nabla ( V + Q ) |_{{\bf x} = {\bf x}(t)} .
\end{equation}

The LR invariant, $I(t)$, of a time-dependent harmonic oscillator
\begin{equation}
\hat{H} (t) = \frac{1}{2 M(t)} {\hat{p}}^2 + \frac{1}{2} M(t) \omega^{2} (t) {\hat{q}}^2
\end{equation}
can be written as (hereafter we adopt $\hbar=1$ units.)
\begin{equation}
\hat{I} (t) = \omega_{I} \left[ \hat{b}^{\dagger} (t) \hat{b} (t) + \frac{1}{2} \right] .
\label{Ib}
\end{equation}
Here we have used the creation and annihilation operators introduced in 
Ref.~\cite{Heis}. Noting that
\begin{equation}
f(t) = \sqrt{ \frac{{g_{-} (t)}}{2 \omega_{I}} } e^{ - i \Theta (t,t_{0} )}
\end{equation}
with $\Theta(t, t_{0} ) =
\int_{t_{0}}^t dt \frac{\omega_{I}}{ M(t) g_{-} (t)} $
satisfies the classical equation of motion:
\begin{equation}
\frac{d}{dt} [M(t) \frac{d}{dt} f (t)] 
+ M(t) \omega^{2} (t) f (t) = 0 .
\label{EOM}
\end{equation}
and using 
\begin{equation}
\frac{d}{dt}  \left[ \sqrt{ \frac{{g_{-} (t)}}{2 \omega_{I}} } e^{- i \Theta (t,t_{0} )} \right]
= - \frac{{i}}{M} \left[ \sqrt{\frac{\omega_{I}}{2 g_{-} (t)}} 
- i \frac{{g_{0} (t)}}{\sqrt{2 \omega_{I} g_{-} (t)}} \right] 
e^{ - i \Theta (t,t_{0} )}
\end{equation}
the operators $\hat{b} $ and $\hat{b}^{\dagger} $ in $\hat{b} (t) = b e^{-i \Theta (t, t_{0} )} $ and $b^{\dagger} (t) = b^{\dagger} e^{i \Theta (t,t_{0} )} $ can be written as 
\begin{eqnarray}
\hat{b} &=& i
[ - M(t) \dot{f}^* (t) \hat{q} (t) + f^{*} (t) \hat{p} (t) ]
\nonumber \\
\hat{b}^{\dagger} &=& - i
[ - M(t) \dot{f} (t) \hat{q} (t) + f (t) \hat{p} (t) ]
\label{bb}
\end{eqnarray}
where $\Theta(t, t_{0} ) = 
\int_{t_{0}}^t \frac{dt}{ 2 M(t) |f(t)|^{2}} $ and 
we normalize $f(t)$ so that 
$[b, b^{\dagger} ] = i M(t) [\dot{f} f^{*} - f \dot{f}^{*} ](t) = 1 $.
(For this condition, the Wronskian of $f$ and $f^*$ should not vanish, i.e. 
they are linearly independent.) 
There are uncountable number of invariants since we have freedom in 
choosing the classical solution $f(t)$: Any linear combination of two 
independent classical solutions is also a solution, or equivalently, we are 
free to choose the parameters $c_1$, $c_2$, and $c_3$ with fixed two 
independent solutions $f_1(t)$ and $f_2(t)$ in Eq.~(7) of Ref.~\cite{Heis}. 
Hereafter when we construct the LR invariant, we choose the function $f(t)$ 
and $\omega_{I} $ instead of choosing 
$c_{1,2,3}$. Then the auxiliary function is $g_{-} (t) = 2 \omega_{I} |f(t) |^{2}  $ and 
the annihilation operator and the eigenstates of the invarint are uniquely 
specified only by $f(t)$. Thus we denote them by $b_f$ and $\left| n \right>_{f}  $, respectively. With 
$f$ fixed, the invariants, $I_f$, may differ by a constant factor $\omega_{I} $, which is irrelevant to the 
construction of quantum states. 

Let us take another classical solution, $F(t)$, which are related by $f(t)$ as
\begin{eqnarray}
F(t) &=& u f(t) + v^{*} f^{*} (t) \\
F^{*} (t) &=& v f(t) + u^{*} f^{*} (t)
\end{eqnarray}
to construct another creation and annihilation operators
\begin{eqnarray}
\hat{b}_{F} &=& i
[ - M(t) \dot{F}^* (t) \hat{q} (t) + F^{*} (t) \hat{p} (t) ]
\nonumber \\
\hat{b}_{F}^{\dagger} &=& - i
[ - M(t) \dot{F} (t) \hat{q} (t) + F (t) \hat{p} (t) ]
\label{b:F}
\end{eqnarray}
Then the relation between two sets of creation and annihilation operators 
is given by the following Bogoliubov transformation or unitary (similarity) transformation:
\begin{equation}
\hat{b}_{F} = u \hat{b}_{f} + v^{*} \hat{b}_{f}^{\dagger} = 
\hat{S}^{\dagger} \hat{b}_{f} \hat{S}
\label{Bb}
\end{equation}
generated by the squeezing operator
\begin{equation}
\hat{S} 
= e^{i \theta_{u} \hat{b}_{f}^{\dagger} \hat{b}_{f}} 
\exp \left( 
\frac{\sigma}{2} e^{+i(\theta_{v} - \theta_{u} )} \hat{b}_{f}^{\dagger 2}
- \frac{\sigma}{2} e^{-i (\theta_{v} - \theta_{u} )} \hat{b}_{f}^{2}
\right) 
\label{sqOp}
\end{equation}
where $\sigma = \cosh^{-1} |u| $. Therefore it is verified that the eigenstate of 
$I_F$ is obtained by squeezing the eigenstate of $I_f$ as
\begin{equation}
\left| n \right>_{F} = \hat{S}^{\dagger} \left| n \right>_{f}
\end{equation}
and for the coherent state $\left| \alpha \right>
= e^{- |\alpha|^{2} /2 } \sum_{n} \frac{{\alpha^{n}}}{\sqrt{n!}}
\left| n \right> $
\begin{equation}
\left| \alpha \right>_{F} = 
\hat{S}^{\dagger} \left| \alpha \right>_{f} .
\label{sq:co}
\end{equation}
Since $\left| \alpha \right>_{f} $ can be obtained by displacing $\left| 0 \right>_{f} $, $\left| \alpha \right>_{F} $ is obtained by displacing and squeezing $\left| 0 \right>_{f} , $ i.e. 
\begin{equation}
\left| \alpha \right>_{F} = 
\left| \sigma, \alpha \right>_{f} =
\hat{S}^{\dagger} (\sigma) \hat{D} (\alpha) \left| 0 \right>_{f}
\label{sq:0}
\end{equation} 
with $\hat{D} (\alpha) = e^{\alpha \hat{a}^{\dagger} - \alpha^{*} \hat{a} } $.

Recently, in Ref.~\cite{Exac}, we have found the explicit form of wave 
function for the $n$-th eigenstate of the invariant (\ref{Ib}):
\begin{equation}
\psi_{n} (q,t) = R_{n} (q,t) e^{i S_{n} (q,t)/ \hbar}
\label{nth}
\end{equation}
where
\begin{equation}
R_{n} (q,t) =
\frac{1}{\sqrt{ 2^{n} n!}}
\left(
\frac{\omega_{I}}{\pi g_{-} (t)}
\right)^{\frac{1}{4}}
e^{- \frac{\omega_{I}}{2 g_{-} (t)} q^{2} }
H_{n} \left( \sqrt{\frac{\omega_{I}}{g_{-} (t)} } q \right) . 
\label{amp}
\end{equation}
and
\begin{equation}
S_{n} (q,t) = - \frac{{g_{0} (t)}}{2 g_{-} (t)} q^{2}  
- \left(n + \frac{1}{2} \right)
\Theta(t,t_{0} ) .
\label{pha}
\end{equation}

When the exact form of the wave function is given, it is straightforward to 
find the quantum trajectory by solving the quantum equation of motion 
(\ref{QEOM}). The quantum equation of motion for the $n$-th eigenstate of 
the invariant is 
\begin{equation}
M(t) \dot{q} = - \frac{{ g_{0} (t)}}{g_{-} (t)} q .
\end{equation}
Using $g_{0} = - \frac{M}{2} \dot{g}_{-} $, we find a solution easily
\begin{equation}
q(t) = \frac{{q(t_{0} )}}{\sqrt{g_{-} (t_{0} )}} \sqrt{g_{-} (t)} .
\label{qm:ei}
\end{equation}
As expected from the result of damped oscillator~\cite{Vand}, the quantum 
trajectory is independent of the quantum number $n$ in a general 
time-dependent oscillator.

Now we consider the coherent state of the invariant
\begin{equation}
\Psi_{\alpha} (q, t)
= e^{- |\alpha|^{2} /2 } \sum_{n} \frac{{\alpha^{n}}}{\sqrt{n!}}
\psi_{n} (q,t)
\label{cow}
\end{equation}
Using $e^{-z^{2} + 2 x z } = \sum_{0}^{\infty} \frac{{z^{n}}}{n!} H_{n} (x) $
with 
$z = \frac{1}{\sqrt{2}} \alpha e^{-i \Theta(t,t_{0} )} $ 
and 
$x = \sqrt{\frac{{\omega_{I}}}{g_{-}}} q $,
we have
\begin{equation}
\Psi_{\alpha} (q,t) = 
\left(
\frac{\omega_{I}}{\pi g_{-} (t)}
\right)^{\frac{1}{4}}
e^{
- \frac{{|\alpha|^{2}}}{2} 
- \frac{\omega_{I}}{2 g_{-} (t)} q^{2} 
- i \frac{g_{0}}{2 g_{-}} q^{2}
- \frac{i}{2} \Theta(t,t_{0} )
}
e^{-z^{2} + 2 x z} .
\label{co:wf}
\end{equation}
We rewrite the above wave function as
\begin{equation}
\Psi_{\alpha} (q,t)= R_{\alpha} (q,t) e^{i S_{\alpha} (q,t)/ \hbar}
\end{equation}
where
\begin{equation}
R_{\alpha} (q, t) =
\left(
\frac{\omega_{I}}{\pi g_{-} (t)}
\right)^{\frac{1}{4}}
e^{
- \frac{{|\alpha|^{2}}}{2} 
- \frac{\omega_{I}}{2 g_{-} (t)} q^{2} 
- \frac{\alpha^{2}}{2} \cos [2 \Theta(t,t_{0} )]
+ \alpha q \sqrt{\frac{{2 \omega_{I}}}{g_{-} (t)}} \cos \Theta(t,t_{0} )
}
\end{equation}
and
\begin{equation}
S_{\alpha} (q,t) =
- \frac{g_{0}}{2 g_{-}} q^{2}
- \frac{1}{2} \Theta(t,t_{0} )
+ \frac{\alpha^{2}}{2} \sin[2 \Theta(t,t_{0} )] 
- \alpha q \sqrt{\frac{{2 \omega_{I}}}{g_{-} (t)}} \sin[\Theta(t,t_{0} )] .
\end{equation}
Therefore, from (\ref{QEOM}), we have the quantum equation of motion for 
the coherent state
\begin{equation}
M \dot{q} = - \frac{{g_{0}}}{g_{-}} q 
- \alpha \sqrt{\frac{{2 \omega_{I}}}{g_{-}}}
\sin \Theta
\end{equation}
which can be easily integrated using
\begin{equation}
\frac{d}{dt} \left( \frac{q}{\sqrt{g_{-}}}\right)
= \alpha \sqrt{\frac{2}{\omega_{I}} } \frac{d}{dt} \cos \Theta ,
\end{equation}
then we have the following quantum trajectory 
\begin{equation}
q(t) = \frac{{q(t_{0} )}}{\sqrt{g_{-} (t_{0} )}} \sqrt{g_{-} (t)} 
+ \alpha \sqrt{\frac{{2 g_{-} (t)}}{\omega_{I}}} 
[ \cos \Theta(t, t_{0} ) - 1 ] .
\label{qm:co}
\end{equation}
One should note that the coherent state (\ref{cow}) of an invariant can be 
interpreted as the displaced squeezed state of another invariant [see 
(\ref{sq:0}) and related remarks]. 
Thus the quantum motion we have found can be, in general, gives a whole 
description of the motion for a squeezed state.

As a simple example we consider the time-independent case with $M(t) = m $ and $\omega(t) = \omega_{0} $. If 
we choose the classical solution 
$f (t) = ( 1 / \sqrt{2 m \omega_{0} }) e^{ - i \omega_{0} t} $ 
and $\omega_{I} = \omega_{0} $ we have
\begin{equation}
\hat{I}_{f} (t) = \hat{H} = 
\frac{{\hat{p}}^2}{2m} + \frac{1}{2} m \omega_{0}^{2} {\hat{q}}^2 .
\end{equation}
For an eigenstate of this invariant, it follows from $g_{-} (t) = 1/m $ and (\ref{qm:ei}) that
\begin{equation}
q(t) = q(t_{0} ) ,
\end{equation}
and for a coherent state, from (\ref{qm:co})
\begin{equation}
q(t) = {q( 0 )}  
+ \alpha \sqrt{\frac{2}{m \omega_{0}}} 
( \cos \omega_{0} t - 1 )  ,
\end{equation}
which are well known results~\cite{Holland}.

While if we choose the classical solutions as 
$F(t) = (\cosh \sigma) f (t) + (\sinh \sigma) f^{*} (t) , $ 
we have
\begin{equation}
\hat{I}_{F} = g_{-} \frac{{\hat{p}}^2}{2} 
+ g_{0} \frac{{ \hat{p} \hat{q} + \hat{q} \hat{p} }}{2}
+ g_{+} \frac{{\hat{q}}^2}{2}
\end{equation}
where
\begin{eqnarray}
g_{-} (t) &=& ( 1 / m ) 
( e^{2 \sigma} \cos^{2} \omega_{0} t 
+ e^{-2 \sigma} \sin^{2} \omega_{0} t), 
\nonumber \\
g_{0} ~(t) &=& - (e^{2 \sigma} 
- e^{-2 \sigma} ) \omega_{0} \cos \omega_{0} t \sin \omega_{0} t , \\
g_{+} (t) &=& m \omega_{0}^{2} 
( e^{2 \sigma} \sin^{2} \omega_{0} t 
+ e^{-2 \sigma} \cos^{2} \omega_{0} t) . 
\nonumber
\end{eqnarray} 
and we have a quantum trajectory, for an eigenstate $\left| n \right>_{F} $
\begin{equation}
q(t) =  q( 0 ) \sqrt{ \cos^{2} \omega_{0} t + e^{-4 \sigma} \sin^{2} \omega_{0} t },
\end{equation}
and for a coherent state $\left| \alpha \right>_{F} $
\begin{equation}
q(t) =  \sqrt{ \cos^{2} \omega_{0} t + e^{-4 \sigma }  \sin^{2} \omega_{0} t }
\left[ q( 0 ) + \alpha e^{\sigma} \sqrt{2  / {m \omega_{0}} } 
( \cos \Theta - 1 )
\right] ,
\end{equation}
with 
$\Theta(t) = \tan^{-1} [ e^{-2 \sigma} \tan (\omega_{0} t - r \pi) ] + r \pi ,~
{\rm for} ~ | \omega_{0} t - r \pi | \leq \pi/2 ~
(r = 0, \pm 1, \pm 2 , ...) . $
As stated before, $\left| \alpha \right>_{F} $ is a displaced 
squeezed state which is obtained by displacing and squeezing the ground 
state of the time-independent harmonic oscillator. Thus we have found the 
quantum motion for the displaced squeezed state.

For comparison with the result of Ref.~\cite{Vand}, we take the damped 
oscillator as another example: $M(t) = m e^{2 \gamma t} $, $\omega(t) = \omega_{0} $ 
with $\omega_{0} > \gamma $. The 
classical equation of motion is
\begin{equation}
\ddot{f} + 2 \gamma \dot{f} + \omega_{0}^{2} f = 0
\end{equation}
and if we choose $f (t) = (1/\sqrt{2 m \Omega}) e^{-\gamma t - i \Omega t} $ with $\Omega = \sqrt{\omega_{0}^{2} - \gamma^{2}} $ as a solution and $\omega_{I} = \Omega $, we have $g_{-} (t) = (1/m) e^{- 2 \gamma t} $ and the quantum trajectory for $\left| n \right>_{f} $ is
\begin{equation}
q(t) = q(0) e^{-\gamma t}
\end{equation}
which is the same result obtained in Ref.~\cite{Vand}. 
However if we choose 
$F(t) = (\cosh \sigma) f(t) + (\sinh \sigma) f^{*} (t) $
we have 
$g_{-} (t) = ( e^{-2 \gamma t} / m ) 
( e^{2 \sigma} \cos^{2} \Omega t 
+ e^{- 2 \sigma} \sin^{2} \Omega t),  $ 
and the quantum trajectory for 
$\left| n \right>_{F} $
\begin{equation}
q (t) = q(0) e^{-2 \gamma t} 
( \cos^{2} \Omega t 
+ e^{- 4 \sigma} \sin^{2} \Omega t). 
\end{equation}
This quantum trajectory is clearly oscillatory while damping. 

In summary, we have constructed LR invariants introducing the creation and 
the annihilation operators which are specified by a classical solution. We 
have found the quantum trajectory for the eigenstates and for the coherent 
state of this invariant. As in the cases of LR invariant, the time evolution 
of quantum operator $\hat{q} (t) $ 
and $\hat{p} (t) $ 
in the Heisenberg picture, and the wave function of the invariant, the 
quantum trajectories are also expressed only in terms of the classical 
solution of the oscillator. The solutions were verified in the 
time-independent oscillator and a damped oscillator.

The quantum trajectory we have found can be easily applied to other 
time-dependent oscillators as far as the classical solutions are known. We 
expect that the oscillator's trajectory give a good description, as an 
example, of explicitly time-dependent system in the de Broglie-Bohm quantum 
theory. Our formalism can also be applied to a Paul trap~\cite{Paul90} or a 
modified Paul trap~\cite{Park} to study the quantum motion of a particle.
\\

This work was supported by the Center for Theoretical Physics (S.N.U.) and 
the Basic Science Research Institute Program, Ministry of Education Project 
No. BSRI-96-2418. One of us (JYJ) is supported by Ministry of Education for 
the post-doctorial fellowship.

\end{document}